\documentclass{camera}

\begin{document}

%
\title{Interpreting the high energy emission of Fermi GRBs}

%
\author{Yi-Zhong Fan}

%
\organization{Niels Bohr International Academy, Niels Bohr Institute,
University of Copenhagen, Blegdamsvej 17, DK-2100
Copenhagen, Denmark}
\maketitle

\begin{abstract}
The high energy emission from Gamma-ray Bursts has some interesting features, including
the absence of the GeV excess in the prompt spectrum, the delayed onset of the GeV emission, and the longer duration of the GeV emission than the prompt soft gamma-ray emission. We suggest that the non-detection of a GeV excess in most GRB spectrum may favor the
magnetic fireball model and the early prompt emission may be dominated by
the photosphere radiation of the breakout material and is thus very soft. The synchrotron
radiation in GeV band can be the dominant component of the high energy afterglow emission, as
speculated in GRB 080319B and then confirmed in GRB 080916C and GRB 090510. A simple estimate
of the thermal radiation of the breakout material has been presented.
\end{abstract}

%
\section{Introduction} Gamma-ray bursts (GRBs) are the most luminous explosions in the
Universe. Though their cosmological origin as well as the
relativistic movement has been firmly established, the radiation
mechanism and the outflow composition are still uncertain. It is
widely believed that the high energy
emission of GRBs can shed light on these two fundamental issues
\cite{FP08}. For example, a distinct GeV-TeV
spectrum excess can be taken as an indication evidence of a baryonic
outflow and a radiation process in addition to synchrotron will be needed, while the absence of
such a component in most spectra may favor the magnetic outflow
model. The Large Area Telescope (LAT) and the
Gamma-ray Burst Monitor (GBM) onboard Fermi satellite
(http://fermi.gsfc.nasa.gov/) can measure the spectrum in a very
wide energy band (from 8 keV to more than 300 GeV), with which some
models may be distinguished.
Since the launch of Fermi satellite on 11 June 2008, significant
detection of prompt GeV emission from GRBs has been
reported in about 11 GRB (see Tab.\ref{tab:sum} and \cite{Abdo09a,Abdo09b,Abdo09c,Fan09a}). The detection rate
is about one burst per month, consistent with the prediction based on the assumption that
the GRB spectrum is a featureless Band function. Indeed, so far the GeV excess, widely expected in the
model calculation, has only been detected in 2 (possibly 3) bursts.
Therefore {\it the absence of a GeV spectrum excess in most GRBs is a well-established fact}. The other two common features of the high energy emission
are {\it the delayed onset} and {\it the long duration of the high energy emission}, respectively. Below we interpret these features.

\begin{table}
\caption{A brief summary of the GRBs with GeV emission (detected by Fermi satellite from 10
August 2008 to 21 Oct. 2009). The data are taken from
http://gcn.gsfc.nasa.gov/gcn3$_{-}$archive.html (see also the talk
by Nicola Omodei in this conference).}
\begin{tabular}{l|c|c|c|c}
\hline GRB & Type & GeV excess & delayed onset & GeV afterglow
\\ \hline
080825C & long & No & Yes &  Yes
\\ \hline
080916C & long & No & Yes &  Yes
\\ \hline
081024B & short & No & Yes & Yes
\\ \hline
090217 & long & No & &  No
\\ \hline
090323 & long & No & Yes &  Yes
\\ \hline
090328 & long & No & & Yes
\\ \hline
090510 & short & Yes & Yes  & Yes
\\ \hline
090626 & long & No & & Yes
\\ \hline
090902B & long & Yes & Yes & Yes
\\ \hline
090926A & long  & In the brightest phase & & Yes
\\ \hline
091003A & long & No & & Yes
\\ \hline
\end{tabular}
\label{tab:sum}
\end{table}

\section{The absence of the GeV spectrum excess} In the standard fireball model the GRB outflow
is baryonic \cite{Piran04}. A GeV spectrum excess can be produced in two ways. (1) In the
standard internal shock model the prompt soft gamma-rays are the synchrotron radiation of the shocked
electrons. The synchrotron self-Compton (SSC) radiation of these electrons will peak in GeV-TeV energy range
and give rise to a GeV excess \cite{Pilla98}. (2) In the photosphere-internal shock model, the
photosphere radiation gives rise to a MeV peak with soft high energy spectrum. The inverse Compton (IC) radiation of the electrons accelerated in the subsequent internal shocks can produce a GeV excess \cite{mr00}. Therefore,
the detection of a distinct GeV excess in the prompt spectrum, for example in GRB 090510 and GRB 090902B, can be taken as a piece of evidence of the baryonic outflow model \cite{Gao09}. The non-detection of such a component
 in most other events does not necessary mean that the baryonic outflow model has been ruled out. For example, one can argue that the SSC of the internal shocks is within the extreme Klein-Nishina regime and is thus very inefficient. As a result the GeV emission is very weak, consistent with the data \cite{Fan09a}. In some  photosphere models, no GeV excess is expected either. The magnetic fireball model \cite{Usov92}, nevertheless, provides us a compelling interpretation. The strong magnetic field in the emitting region can suppress the inverse Compton radiation of electrons accelerated in the magnetic dissipation process effectively, in agreement with the data.

Among the possibilities outlined above, we think {\it the magnetic fireball model is favored} in light of the following other facts (see \cite{Fan09b}
for a summary and for the references): (a) The analysis of some well-studied optical flashes of GRBs reveal that the magnetic fields in
 the reverse-shock region are much stronger than that in the forward-shock region, so that
  the GRB outflows are probably magnetized. (b) The absence of a distinct
  thermal spectrum component in most GRBs is consistent with the Poynting-flux dominated outflow model. (c) The non-detection of bright optical flash in most GRB afterglows can be attributed to
a mild or high magnetization of the outflow. 
(d) The (possible) detection of the high linear polarization degree of some GRBs suggests that the magnetic field
involved in the synchrotron radiation could be globally ordered.

\section{The delayed onset of the GeV emission} Various possibilities have been proposed in the literature. A kind of model is the different physical origin of the GeV and MeV photons. For example, the MeV photons
may have an internal origin while the GeV photons may have an external/outer origin. Or alternatively the MeV
and GeV photons are the radiation of electrons and protons respectively and the onset delay reflects the
longer acceleration timescale of the protons than the electrons.
Limited by the space here we focus on the interpretation based on some
plausible physical processes taking place at the early stage of the GRBs \cite{Fan09a}: (i) In both the collapsar and the compact star merger models for GRBs, the early outflow likely
suffers more serious baryon pollution and thus has a smaller
Lorentz factor than the late ejecta \cite{ZhangW04}. Hence the emitting region is optically thick for
the GeV photons.
(ii) In the collapsar scenario, before the breakout, the initial outflow
is choked by the envelope material of the massive star and is very hot \cite{ZhangW04}.
{\it The emission
of the breakout material is likely dominated by the
thermal component from the photosphere and may last a few seconds,
as shown below.} Following \cite{ZhangW04,Lazz09}, we assume that the progenitor has a size $R_* \sim 10^{11}$ cm. The breakout outflow is only mildly relativistic (i.e., the bulk Lorentz factor $\Gamma_* \sim 10$) and is radiation-dominated
($e'/n'm_{\rm p}c^{2} \sim 50$, where $e'$ ($n'$) is the comoving thermal energy (number) density of the outflow)  with an isotropic-equivalent luminosity
$L_* \sim 10^{53}~{\rm erg/s}$. With the simplified assumption that the ejecta has no sideways expansion,
the outflow acceleration is described by the well-known relation $\Gamma \sim \Gamma_* R/R_*$ \cite{Piran93,Mesz93}.
On the other hand, the outflow becomes transparent at a radius
$R_{\rm ph} \sim 1.2\times 10^{13}L_{*,53}\Gamma_{2}^{-2}\eta_{2.7}^{-1}$ \cite{Paczyn90},
where $\eta=\Gamma_*e'/n'm_{\rm p}c^{2}$, and the convention
$Q_x=Q/10^x$ has been adopted in cgs units. At $R\sim R_{\rm ph}$, the bulk Lorentz factor of the outflow is
\begin{equation}
\Gamma_{\rm ph} \sim 230~L_{*,53}^{1/3}\Gamma_{*,1}^{1/3}R_{*,11}^{-1/3}\eta_{2.7}^{-1/3},
\end{equation}
which is smaller than $\eta$ as long as $\eta>280L_{*,53}^{1/4}\Gamma_{*,1}^{1/4}R_{*,11}^{-1/4}$.
If so, the outflow is transparent and the thermal photons can escape freely. The thermal radiation efficiency can be estimated as
\begin{equation}
\epsilon_{\rm th} \sim (\eta-\Gamma_{\rm ph})/\eta \sim 55\%
\end{equation}
for the typical value adopted. The observed temperature and duration of the photosphere radiation can be estimated as
\begin{equation}
T_{\rm obs} \sim T_* \sim 100~{\rm keV}~(1+z)^{-1}L_{*,53}^{1/4}R_{*,11}^{-1/2}\Gamma_{*,1}^{1/2},~~~~T_{\rm dur}\sim (1+z)R_*/c.
\end{equation}
{\it The above crude estimate provides us the guideline to understand the numerical simulation results reported in} \cite{Lazz09}.

The outflow launched after the
breakout of the early ejecta can escape from the progenitor freely.
The consequent energy dissipation can be strong enough to produce
energetic non-thermal emission and to outshine the simultaneous
photosphere radiation. We then expect a soft to hard spectrum evolution, as revealed in the data.

\begin{figure}
\begin{picture}(0,180)
\put(0,0){\includegraphics{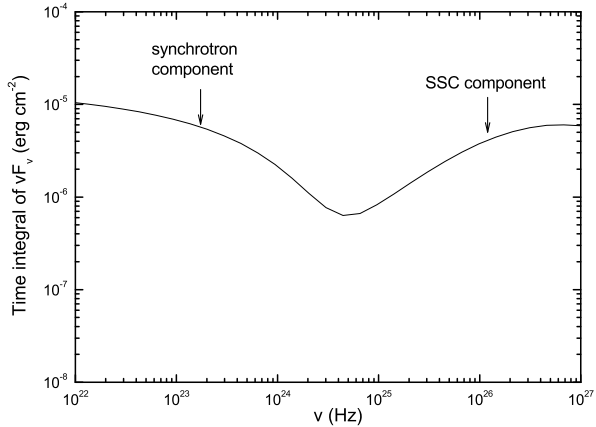}}
\end{picture}
\caption{Expected high energy forward shock emission of GRB 080319B, including the synchrotron +
SSC components, in the time interval 60 sec $-$ 2000 sec \cite{Zou09}.} \label{fig:Zou-09}
\end{figure}

\section{The GeV afterglow emission} The spectrum of the
synchrotron radiation of shocked electrons can
extend to an energy $\sim 30 {\cal A}\Gamma_{\rm i}/(1+z)~{\rm MeV}$
\cite{Chengwei96}, where $\Gamma_{\rm i}$ is the bulk Lorentz
factor of the emitting region and ${\cal A} \sim (1,~2\pi)$,
depending on the comoving acceleration timescale of the particles.
But usually the IC scattering plays a more important role in
producing high energy afterglow emission (see \cite{FP08} and the references therein).
The situation changed in
GRB 080319B, the naked-eye burst with abundant optical and X-ray
afterglow data. With the well constrained parameters, Zou et al.
\cite{Zou09} have shown that {\it the forward shock synchrotron
radiation dominates over the SSC radiation up
to an energy $\sim 10$ GeV} (see Fig.\ref{fig:Zou-09}). The detection prospect for LAT is pretty
good. The forward shock parameters of GRB 080916C seem rather
similar to those of GRB 080319B \cite{Gao09}. A strong forward shock synchrotron
GeV emission is naturally expected and may be able to account for the observational data \cite{Kumar09}.
The same conclusion can be drawn for GRB 090510 \cite{Gao09,Ghir09}.

If the high energy afterglow is due to the IC radiation of the
forward shock electrons, there is a simple method to estimate the
number of seed photons, regardless of their origin
(either the late prompt emission from the central engine or the
synchrotron radiation of the forward shock electrons). Following
\cite{Fan06}, the possibility of one seed photon being scattered
in the forward shock region can be
estimated as $\tau_{\rm ISM} \sim 4.2\times
10^{-8}~E_{\rm k,53}^{1/4}n_0^{3/4}t_3^{1/4}[(1+z)/2]^{-1/4}$ and
$\tau_{\rm wind} \sim 7.3 \times
10^{-6}~A_*^{3/2}E_{\rm k,53}^{-1/2}t_3^{-1/2}[(1+z)/2]^{1/2}$,
respectively, where $E_{\rm k}$ is the isotropic-equivalent kinetic energy of the outflow, $n$
  is the number density of the ISM and $A_*$ is the wind parameter.

With the given time interval $\Delta t$ and the time-averaged high energy photon flux  ${\rm F}_{\rm >100 MeV}$,
 it is straightforward
to estimate the (isotropic-equivalent) number
of total seed photons  \cite{Gao09}
\begin{equation}
N_{\rm seed} \sim {4\pi D_{\rm L}^2  \over (1+z)^{2}}  {{\rm
F}_{\rm >100MeV}  \over \tau}\Delta t,
\end{equation}
with which we can see wether the IC radiation origin of the GeV afterglow is reasonable or not,
where $D_{\rm L}$ is the luminosity distance of the burst to us.
The answer seems negative for GRB 080916C and GRB 090510 since the required $N_{\rm seed} \sim 10^{65}$
 is too big to be realistic \cite{Gao09}.

\section{Discussion} After about one year's successful performance of the Fermi satellite,
the high energy emission properties of GRBs are much better understood than ever before.
The prompt high energy spectrum can be an ideal probe of the very early GRB physics.
For instance, the delayed onset of the GeV emission may reflect
the physical condition of the early outflow. There are two (or more) possibilities.  One is that
 the early outflow suffers serious baryon pollution. The
Lorentz factor is so small that the emitting region is optically thick for the GeV photons. The other is that in the collapsar scenario the early outflow consisting of the breakout material becomes transparent before getting effectively accelerated and then loses its energy mainly via thermal radiation. The spectrum of such a transient is expected to be quasi-thermal and the duration is about a few seconds, consistent with the early emission properties of some GRBs.

As for the absence of the GeV excess in the prompt spectrum of most GRBs, the baryonic outflow model
 has not been ruled out. But together with some other data, including the non-detection of
 the thermal spectrum component and the bright optical flashes in most GRBs, the weakly magnetized reverse shock region found in almost all optical flash modeling, and the moderate/high linear polarization detected in the optical flash of GRB 090102 and possibly in the prompt $\gamma$-ray emission of a few GRBs, the magnetic outflow model may be favored. {\it One prediction of such a model is the weak high energy neutrino emission from GRBs} unless the magnetic energy has been effectively dissipated before the prompt emission phase.

The origin of the GeV afterglow is not well understood. It was widely believed that the IC
radiation of the forward shock will be the dominant component of the high energy afterglow. However,
for GRB 080916C and GRB 090510, the synchrotron radiation of the forward shock may account for the data, confirming the early speculation made in GRB 080319B. {\it Please bear in mind that} \cite{Fan06b,Gao09}: In modeling the optical and X-ray afterglow emission, the IC cooling usually should be taken into account; While in estimating GeV synchrotron emission, the IC cooling is likely effectively suppressed by Klein-Nishina effect and can be ignored. A small GeV ``excess" in the afterglow spectrum can be produced in this way. For the afterglow photons above 10 GeV, as detected in GRB 940217 \cite{Hurley94} and GRB 090902B \cite{Abdo09c}, an IC origin is likely.

{\bf Acknowledgments.} {YZF thanks the conference
organizers for financial support, T. Piran, Y. C. Zou, W. H. Gao
and D. Xu for fruitful collaborations,  and B. Zhang, K. Toma and X. F. Wu for helpful communications.}



%
\end{document}